\documentclass[times,3p,final]{elsarticle}

\usepackage{graphicx}
\usepackage{amssymb}
\usepackage{amsmath}
\usepackage{epsf}

\journal{Physica A}

\begin{document}

\begin{frontmatter}

\title{Income distribution patterns from a complete social security database}

\address[label1]{Department of Theoretical and Computational Physics, Babe\c{s}-Bolyai University \\ Kog\^{a}lniceanu street 1, RO-400080, Cluj-Napoca, Romania}
\address[label2]{Interdisciplinary Computer Modelling Group, Hungarian University Federation of Cluj \\ Baba Novac street 23/2, RO-400080, Cluj-Napoca, Romania}
\address[label3]{Departamento de F\'{\i}sica e Astronomia and Centro de F\'{\i}sica do Porto, Faculdade de Ci\^{e}ncias, Universidade do Porto,\\
     Rua de Campo Alegre s/n, 4169-007 Porto, Portugal}
\cortext[cor1]{Corresponding author.}

\author[label1]{N. Derzsy}
\author[label1,label2]{Z. N\'{e}da\corref{cor1}}\ead{zneda@phys.ubbcluj.ro}
\author[label3]{M. A. Santos}

\begin{abstract}
We analyze the income distribution of employees for  9 consecutive years (2001-2009) using a complete social security database for an economically important district of Romania.
The database contains detailed information on more than half million taxpayers, including their monthly salaries from all employers where they worked.
Besides studying the characteristic distribution functions in the high and low/medium income limits, the database allows us a detailed dynamical study by following the time-evolution of the taxpayers income. To our knowledge, this is the first extensive study of this kind (a previous japanese taxpayers survey was limited to two years).
In the high income limit we prove once again the validity of Pareto's law, obtaining a perfect scaling on four orders of magnitude in the rank for all the studied years.
The obtained Pareto exponents are quite stable with values around  $\alpha \approx 2.5$, in spite of the fact that during this period the economy developed rapidly and also
a financial-economic crisis hit Romania in 2007-2008. For the low and medium income category we confirmed the exponential-type income distribution.
Following the income of employees in time, we have found that the top limit of the income distribution is a highly dynamical region with strong fluctuations in the rank.
In this region, the observed dynamics is consistent with a multiplicative random growth hypothesis.  Contrarily with previous results
obtained for the japanese employees, we find that the logarithmic growth-rate is not independent of the income.
\end{abstract}


\end{frontmatter}

\section{Introduction}

At the end of the twentieth century the research in many modern fields of sciences condensed around complex systems. Statistical physicists got involved in such topics of
interdisciplinary research, applying classical statistical physics methods and models to understand the structure and evolution of such systems.
Many physicists chose to study systems of economic nature, like trade relations, economic transactions, wealth distributions, company interdependencies, etc.
These researches led to the development of econophysics, which became a modern interdisciplinary field \cite{econophysics} of statistical physics.
In the present period of global financial and economic turmoil, this topic is of great interest for everyone who is preoccupied with understanding the complex behavior of
our society and wants to speculate about the future.
One interesting and much debated problem in the field of econophysics, is to understand and model the wealth and income distribution in our society.
With this aim, many data have been collected and many models have been elaborated. An excellent overview of the current standing in this field is given in the
 review paper of Yakovenko and Rosser \cite{yakovenko_2009}. Our research intends to contribute to this field by analyzing
a huge and complete social security database from Romania. The data set contains monthly information regarding employers with head-office registered in the Cluj district,
their employees and the paid salaries for years between 2001 and 2009. This codified social security dataset offers unique possibility to
study the individual income distribution and dynamics on a relatively long period. The aim of the present study is to investigate the dynamics of the
income for the top taxpayers, and to find experimental evidences that would help future modeling attempts for the quite general
Pareto-type distribution of their wealth.

Vilfredo Pareto, an italian economist, observed that the wealth distribution in societies obeys a general law, that later became known as Pareto's law in honor of his work \cite{pareto_1897}.
His observation states that the cumulative income or wealth distribution for the richest $3-5\%$ part of  several countries and cities in the XV-XIX
century Europe follows a power-law. The (symmetric of the) exponent of this power-law is denoted by $\alpha$ and it is named the Pareto exponent. Nowadays, when most of the financial and economic data
are available in electronic format, many recent studies confirmed his predictions with a much improved statistics.  Personal income and wealth distribution studies coming from India \cite{sinha_2006}, Japan \cite{ishikawa_2005,ishikawa_2006}, UK \cite{dragulescu_2001a}, continental Europe \cite{clementi_2007}, USA \cite{kitov_2005,silva_2005} and Brazil \cite{figueira_2011} support this conjecture.
Indirect wealth data originating from history, such as ancient Egypt \cite{abul-magd_2002} or a medieval society \cite{hegyi_2007}, also confirm the universality of the Pareto's law.
Experimental data revealed also that in the limit of low and medium wealth/income region the shape of the cumulative distribution is
well-fitted by either an exponential or a log-normal function.

Measuring the wealth is a problematic task, since it includes several types of incomes and properties and, more importantly, these data are of highly confidential nature.
Thus studying wealth-distribution is more difficult and often leads to inaccurate results caused by incomplete databases.
Although wealth and income are related and both of them show the Pareto's principle, the distribution of the two quantities have clearly separable Pareto exponents.
The distribution of wealth is usually broader than the distribution of income, or equivalently, the Pareto index for wealth distribution is usually
smaller than the corresponding one for income \cite{sinha_2006}. More specifically, the measured $\alpha$ values  for the
individuals income distribution span a quite broad interval, typically in the $1.5-2.8$ range, while studies focusing on
the wealth distribution show a smaller Pareto index, usually in the $0.8 -1.5$ interval \cite{richmond_2005}. This
large variation of $\alpha$ indicates the absence of universal scaling in this problem, a feature which modeling efforts have to take into account and reproduce.

Several theoretical models were created aiming to reproduce the observed distributions.
In case of wealth distribution, the most popular models are agent-based approaches where the wealth of agents varies in a multiplicative and random manner and they can interchange
money following pre-established rules. The interaction between the agents can be either local or global. Such models were successful in reproducing many features
of the wealth-distribution curves \cite{bouchard_2000,solomon_2001,scafetta_2004}.
Asset exchange models  are  also very popular nowadays \cite{grupta_2006,gade_2007,gade_2009}. Trade is the crucial ingredient of these models,
and is taken into consideration by the fact that pairs of randomly chosen agents exchange part of their money while saving the
remaining fraction \cite{chatterjee_2003, chatterjee_2004,pianegonda_2003,sinha_2005}.  For randomly
distributed and quenched saving factors, a Pareto-type wealth distribution with $\alpha=1$ exponent is found \cite{chatterjee_2006}. Variants
of this model considering asymmetric exchanges are able to generate $\alpha<1$ Pareto exponents as well \cite{chatterjee_2004}.
These models are thus able to explain different Pareto index values by varying the free parameters in the wealth-exchange rule.
Due to the complex structure of the underlying social networks on which the wealth-exchange is realized, researchers proposed models
implementing the network approach. In such models the economic interactions between agents take place on small-world or scale-free
network topologies \cite{iglesias_2003,garlaschelli_2004}.  A successful approach in such direction is the model based on first-degree family relations networks
that successfully generates both a realistic wealth distribution and a social network topology \cite{coelho_2005}. For a more complete review
of wealth/income-distribution models we recommend again the review paper of Yakovenko and Rosser \cite{yakovenko_2009}.

The income distribution was also modeled by means of statistical physics approach. The Fokker-Planck equation can be applied for describing the time-evolution of the
income distribution function \cite{silva_2005}. In order to get a stationary solution for the income distribution, one has to postulate how the income changes in time ($\Delta W$)
as a function of the present income value $W$. If for the low income region, it is assumed that $\Delta W$ is independent of $W$ (additive diffusion), while for the top income class one considers
 $\Delta W \propto W$ (multiplicative diffusion), one  gets the right exponential
distribution for the low and medium income region and the power-law distribution for the high income limit. A combination of additive and multiplicative processes has also been studied \cite{yakovenko_2009_springer}; both deterministic and random \cite{milakovic_2005} growth rules have been considered. Multiplicative growth is usually associated with income from bonuses, investments and capital gain; salaries can increase (decrease?) by a constant (merit bonus) or proportinally (cost of living raise, in percentage).

 In order to
have experimental evidence for the $\Delta W$ versus $W$ dependence assumed in theoretical models, exhaustive data for several consecutive years are needed, where one can clearly identify and follow the income of all individuals. This type of information is not easily available --- as far as we know,
such studies have only been performed for two consecutive years in Japan \cite{aoyama_2003,fujiwara_2003}. The Japanese researchers concluded that the distribution
of the growth rate in one year is roughly independent of income in the previous year.  In \cite{aoyama_2003,fujiwara_2003} the authors also argue, that this independency, combined with an approximate time-reversal symmetry, leads to Pareto law, but such claim has been formally proven to be false \cite{bottazzi_2009}.
The dataset available for us, spanning 9 consecutive years and having around half million taxpayers, offers excellent possibilities to reconsider with much better accuracy the $\Delta W$ versus $W$ relationship, and to bring new evidences supporting or disproving the results obtained for Japan.

The rest of the paper is organized as follows. In the first part we briefly  present our database and discuss some features of the income distribution
in the studied geographical region. We study separately the high income limit where the Pareto law is relevant and the low and medium income region where an exponentially
decaying distribution function is  expected. The relatively stable Pareto exponent obtained for all the 9 consecutive years
motivated the second part of the work, where we have collected experimental results on the rank and income variation of individuals. These results would allow
a better foundation for all models that aim to explain the shape and stability of the income distribution function.  With this aim, we have studied how the rank of individuals in the
top income limit changes over time and how their income is related to its previous value over the years.  In this part we also study the $\Delta W$ versus $W$ relationship
considering different time-windows and different representations.

 \section{The used dataset and income distribution}

Our dataset was provided by the Social Security office of the Cluj district in Romania. It contains the personal income of the employees obtained from their salaries at each
employer where he/she worked. The data are on monthly level and span an interval of nine consecutive years in the 2001-2009 time-period.
Each employee is uniquely identified by an encrypted code generated from his/her personal identification number. In a similar manner each employers' name is
also encrypted with a unique number generated from their financial identification number. In such manner the privacy of the employees and employers is not
endangered and the dataset is suitable for research. Since the same employee can have salaries from different employers in the same period, it is  important to sum up workers' salaries from all employers. This summed salary will be referred to as a person's \textit{income}, and the distribution of these quantities will be analyzed. For the sake of an easier
understanding, all incomes are converted to Euros, following the average exchange rate for the given year. The results are presented thus using this currency.

The complete dataset is more than 7 GB and contains information on 535 401 employees and 39 398 employers.
This database offers excellent possibilities for a thorough statistical analysis of the income distribution since this is a complete mapping of an
extended geographical and economic region. The fact that we do have results on 9 consecutive years allows to perform dynamical measurements, following the
employees income evolution.

\begin{figure}[ht!]
  \centering
  \begin{center}
    \includegraphics[scale=0.4]{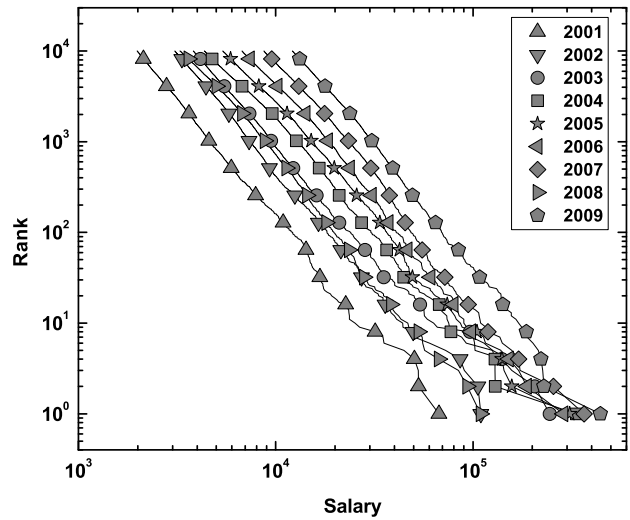}
  \end{center}
  \caption{Rank-salary plot on double-logarithmic scale for the top 10000 salaries (per year nominal in Euro). Data for Cluj district (Romania) for all
  registered workers between years 2001-2009. The exponents and coefficients resulted from the power-law fit are presented in Table \ref{tab1}.}
  \label{fig1}
\end{figure}

The total salaries for each year separately and for the nine consecutive years are computed. From this quantities we have constructed the cumulative distribution function
$P_>(W)$ by a simple rank-income plot.  This is done separately both for the top (which we chose as the 10000 highest incomes) and for the low and medium income category (rank $>10^4$).
In such manner it was possible to analyze the Pareto-type scaling ($P_>(W)=CW^{-\alpha}$) for the high income region and the exponential distribution for the low and medium income category. The presence of the Pareto law is illustrated by using a double-logarithmic plot. Results for each year are displayed on Figure \ref{fig1}. This Figure illustrates nicely that
the power-law scaling holds for 4 orders of magnitude in the rank domain and 2 orders of magnitude in the salary domain. The $\alpha$ Pareto exponent determined from a
a power-law fitting are rather stable (Table \ref{tab1}), although in 2008 the financial crisis hit Romania and the salaries were greatly reduced. This can be observed also in the curves of
Figure \ref{fig1} and the value of the proportionality factor, $C$, given in Table \ref{tab1}. The constant nature of the Pareto exponent is different from what it has been
obtained for USA  for in years 1983-2001, where the Pareto exponent decreased after the stock market crash affected the economy and registered an
increase for periods of financial recovery \cite{silva_2005}. The Pareto exponent value around $2.5$ obtained by us is in agreement with values
obtained in previous income studies \cite{clementi_2007}, \cite{fujiwara_2003}.

\begin{figure}[ht!]
  \centering
  \begin{center}
    \includegraphics[scale=0.4]{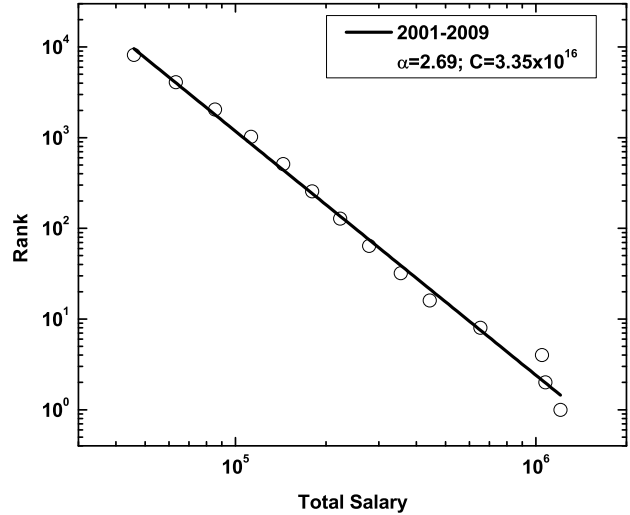}
  \end{center}
  \caption{Rank-salary plot on double-logarithmic scale for the top 10000 total income (in Euro) in the 2001-2009 time-interval.
  The power-law fit is indicated by a thick continuous line. The obtained Pareto exponent, $\alpha$, and proportionality constant $C$ is given in the legend.
  }
  \label{fig2}
\end{figure}

Analyzing the total income of each employee together for the 9 consecutive years, we still find that the Pareto law holds -- Figure \ref{fig2} -- in the top income limit. A power-law fit
suggests that the $\alpha$ Pareto exponent is similar ($\alpha=2.69$) with the ones obtained for each individual year.

\begin{table}[h]
\centering
\begin{tabular}{lccc}
\hline
      & \textbf{$\alpha$} & \textbf{$C$} \\
\hline \hline
\textbf{2001} & $2.54$ & $2.19\times10^{12}$ \\
\textbf{2002} & $2.50$ & $8.47\times10^{12}$ \\
\textbf{2003} & $2.14$ & $3.05\times10^{11}$ \\
\textbf{2004} & $2.33$ & $3.01\times10^{12}$ \\
\textbf{2005} & $2.43$ & $1.27\times10^{13}$ \\
\textbf{2006} & $2.60$ & $1.01\times10^{14}$ \\
\textbf{2007} & $2.61$ & $2.16\times10^{14}$ \\
\textbf{2008} & $2.67$ & $2.95\times10^{13}$ \\
\textbf{2009} & $2.71$ & $1.39\times10^{15}$ \\
\hline
\end{tabular}
\caption{The $\alpha$ Pareto index and $C$ coefficients for years 2001-2009.\label{tab1}}
\end{table}

\begin{figure}[ht!]
  \centering
  \begin{center}
    \includegraphics[scale=0.4]{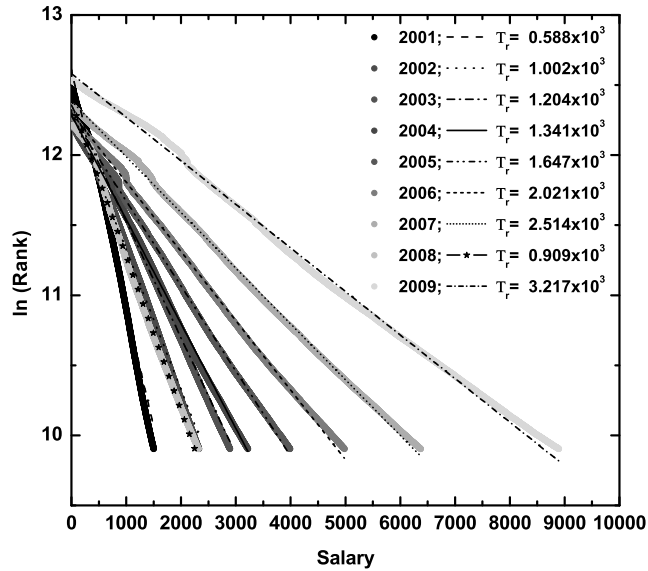}
  \end{center}
  \caption{Rank-salary plots for several years on log-linear representation for employees with rank higher than $2 \times 10^4$.
  The exponential fits $P_>(W)=K \cdot exp(-W/T_{r})$ of the cumulative income distributions are illustrated with lines.
  The $T_r$ income temperature values are given in the legend.}
  \label{fig3}
\end{figure}

In the low and medium income region the income distribution can be fitted by $P_>(W)=K \cdot exp(-W/T_{r})$, where $T_{r}$ is called the income temperature \cite{atkinson_2000,dragulescu_2001}. The rank-salary curves for the employees with rank bigger than $2 \times 10^4$ are shown in
Figure \ref{fig3}. Up to a proportionality constant, these curves are again equivalent to cumulative distribution functions.
 The straight lines in the log-linear representation suggest the validity of the exponential decay. The income temperature values
obtained from the exponential fit are indicated in the legend of Figure \ref{fig3}. The obtained income temperatures present a monotonically
increasing tendency as a function of time, with exception of the austerity year 2008, when the drastic salary cutoffs were implemented.
For this year a sharp decrease in this parameter is observable.

Figures \ref{fig1}-\ref{fig3} suggest that the used data confirm both the validity of the power-law scaling in the high income limit and the validity of the exponential
distribution for the low and medium income region.  Hence we can state that the targeted society presents the same general income characteristics as the
societies investigated in previous studies. This observation empowers us to believe that the dynamics hidden in our data has also universal features, and results obtained from
it can be extrapolated for other societies as well.

\section{Rank dynamics and stability of the Pareto law}

The relatively constant value of the Pareto exponent for the studied 9 year interval is an intriguing fact taking into account that
the income temperature characterizing the exponential income distribution in the low and medium income class changes  constantly
and the $C$ proportionality constant in the $P_>(W)=C \cdot W^{-\alpha}$  power-law fluctuates significantly (see Figure \ref{fig1} and Table \ref{tab1}).

\begin{figure}[ht!]
  \centering
  \begin{center}
    \includegraphics[scale=0.7]{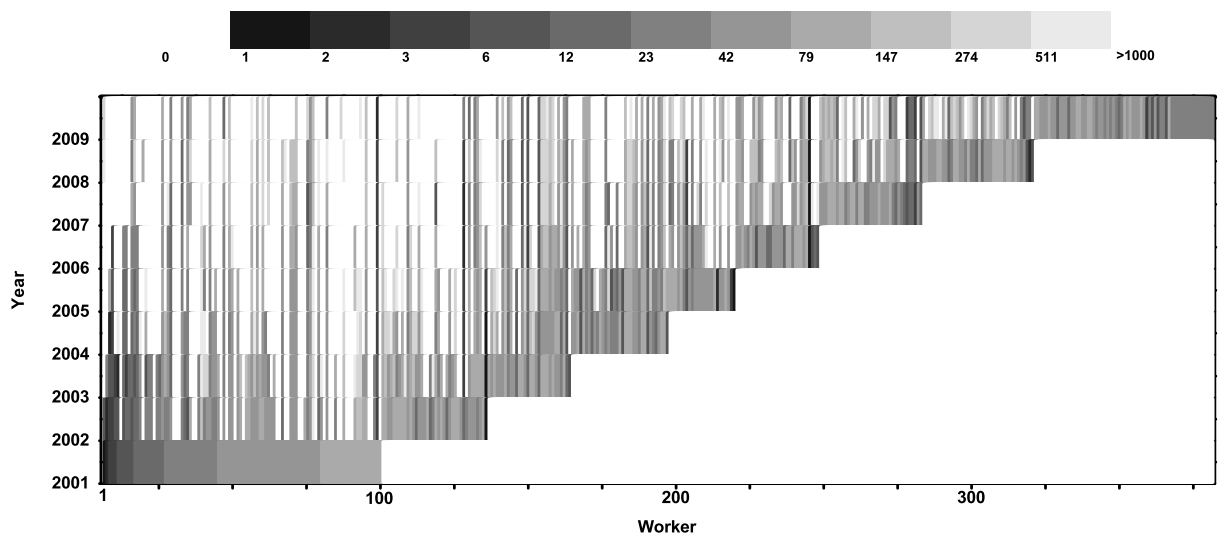}
  \end{center}
  \caption{Rank evolution in time in the top 100 limit. The  horizontal axis is the worker's identification number, while the vertical axis represents the studied year.
  The gray scales highlight the ranks employees have over the years (see the upper legend). Ranks above 1000, as well as years with no income for that particular worker are painted white.}
  \label{fig4}
\end{figure}

In order to understand the relative stability of the Pareto exponent, we decided to follow first the dynamics of the employees' rank
in the top salary limit. A first possible explanation for such a stability would be a relatively stable (in time) rank of individual employees.
To check this hypothesis, in Figure \ref{fig4} we illustrate the workers' rank evolution in time using a gray-scale coding. For the first year (2001) we consider the employees with the
top 100 salaries. For the next year (2002) we include in the list the employees that are breaking-in in the top 100 list. In the same manner for all further years we follow up
the rank of the previously included employees plus the new ones that make it in the top 100. This plot immediately convinces us that the top region of the
income distribution is a highly dynamical one, where constantly new persons appear and disappear. Annually at least $20\%$ of new people break-in the top 100, and
those who are in this top region can fall several orders of magnitude in the rank in two consecutive years. These facts prove that the stability of the income distribution
functions' shape in the high-income limit and the corresponding Pareto exponent are not maintained by the same workers. Employees change their workplaces, positions or jobs,
they may retire or many other conditions may arise that can influence the variation of their salaries, like promotion or demotion.

\begin{figure}[ht!]
  \centering
  \begin{center}
    \includegraphics[scale=0.5]{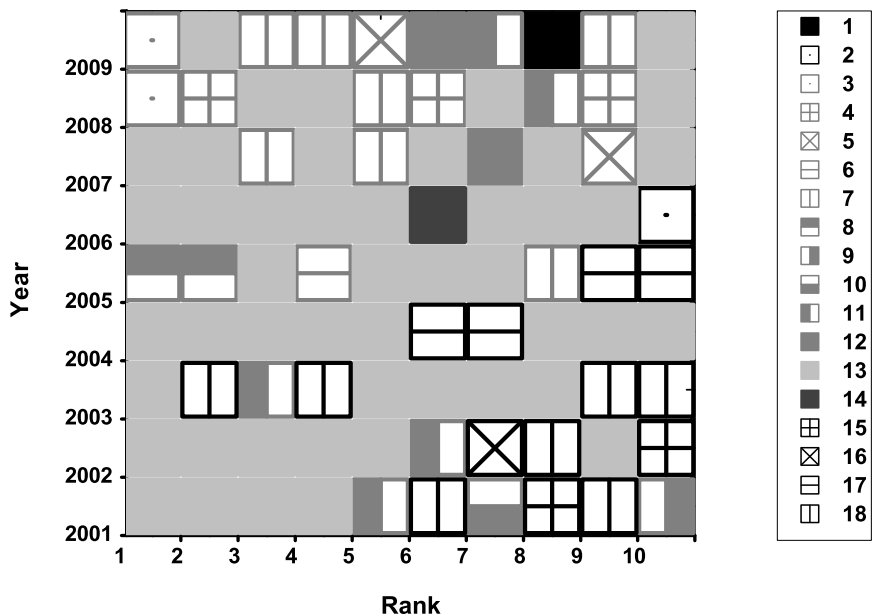}
  \end{center}
  \caption{Employers for the top 10 highly payed position for each year. In this top-list there are in total 18 employers, each of them being coded with different filling patterns.}
  \label{fig5}
\end{figure}

A second possible explanation for the stability of the Pareto's law and exponent value in such a rapidly changing top list would be the following: although the
employees are changing, the jobs are fixed. This means that the highly payed positions are given and employees hired for these positions come and go. In such manner
the employee mobility will not affect the shape of the high income distribution, nor the value of the resulting Pareto exponent. This hypothesis can be verified by selecting in each year
the top jobs and by following the employer who offered these positions. For fixed jobs in the top list their employers have to be also fixed.
In Figure \ref{fig5} we analyze the highly payed top 10 jobs in such aspect. For each year we have identified the employers offering these jobs and
coded them with different filling pattern. The results in Figure \ref{fig5} suggest that, even for the highly payed top 10 offered jobs, the employers are not stable in time, changing from year to year.
In other words, the stability of the top highly payed jobs is also false, and even these jobs (or the employers offering them) are constantly changing.

The results presented in Figure \ref{fig4} and \ref{fig5} imply that the stability of the Pareto law and the corresponding exponent for a given economic region  is not a consequence of a quasi-static rank of the employees or of the available jobs. It is rooted thus somewhere much deeper... and a quite general
statistical explanation should be behind it, as previous modeling efforts had already assumed \cite{yakovenko_2009}.  One should also recall that in the studied period a financial crisis
hit the economy and all salaries were drastically reduced. In spite of this, the Pareto law still remained valid and the Pareto exponent did not change in a significant manner.
The above results are important experimental facts for future modeling efforts, supporting the presently available statistical descriptions based on random growth.

\section{Income dynamics of the rich}

We have argued in the introduction that modeling efforts for income distribution rely on some hypothesis on the income evolution. In the high income (Pareto law) limit they assume a either a multiplicative growth \cite{yakovenko_2009} or the independency of the growth rate from the income. The works of Aoyama et al. \cite{aoyama_2003} and Fujiwara et al. \cite{fujiwara_2003} were the first targeting empirically the income dynamics of the rich in Japan by examining about 80 000 high-income taxpayers for two consecutive years.

Our database offers excellent possibilities for testing, with a more detailed statistics, both the multiplicative growth hypothesis, and the independency of the growth rate as a function of
the income. All the results that are presented here are targeting the high-income category, considering the top $10^4$  taxpayers from the database.

\begin{figure}[ht!]
  \centering
  \begin{center}
    \includegraphics[scale=0.6]{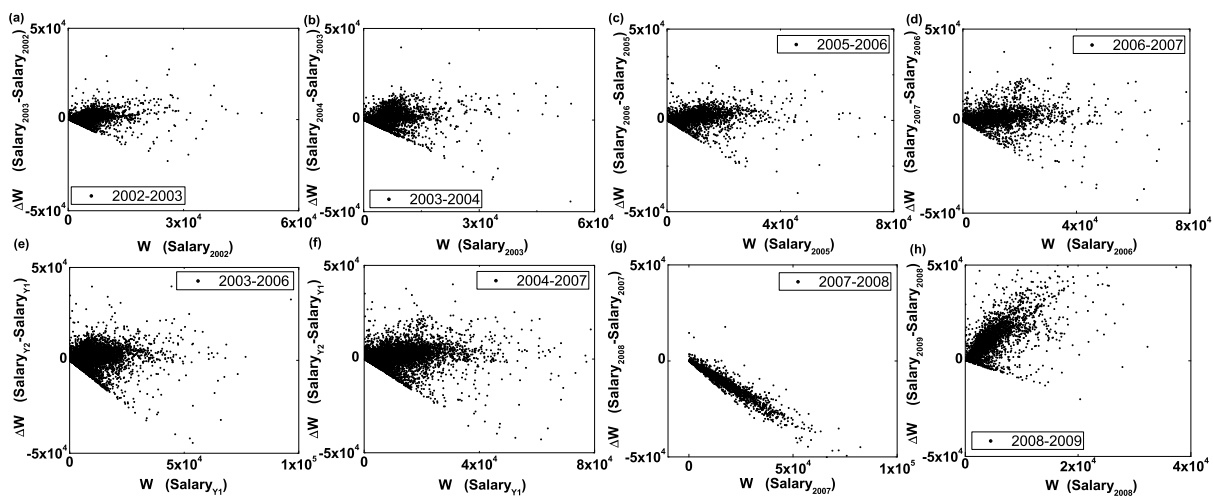}
  \end{center}
  \caption{Time evolution of the income distribution. (a)-(d) The figures plot the salary differences between two consecutive years as a function of the initial salary. (e) and (f) represent data points for 3 years time periods constructed in the same manner. (g) shows a clear decreasing linearity, while for the following year (h) an increasing trend can be observed.}
  \label{fig6}
\end{figure}

First, one can visually test whether there is any relation between the changes in the salaries (between times $t$ and $t+\Delta t$) and the initial salary
 (at time $t$).
One can investigate this by considering several initial $t$ time-moments and several $\Delta t$ time-windows. Some generic results for the
top $10^4$ taxpayers are plotted on Figure \ref{fig6}.
These results indicate that the changes of a given income $W$ fluctuate in a considerable manner. The triangle-like structure
generated by the scattered points suggests that a random multiplicative growth is a good approximation for most of the years. For the austerity years 2007-2008
when the financial-economic crisis hit Romania, the structure is different and we obtain for 2007-2008 a clear linear decreasing trend ($-\Delta W \propto W$),
followed by a linear increase ($\Delta W \propto W$) in the next year. Apart from these special years, the results confirm the hypothesis of models formulated in
\cite{milakovic_2005,yakovenko_2009_springer}.

\begin{figure}[ht!]
  \centering
  \begin{center}
    \includegraphics[scale=0.6]{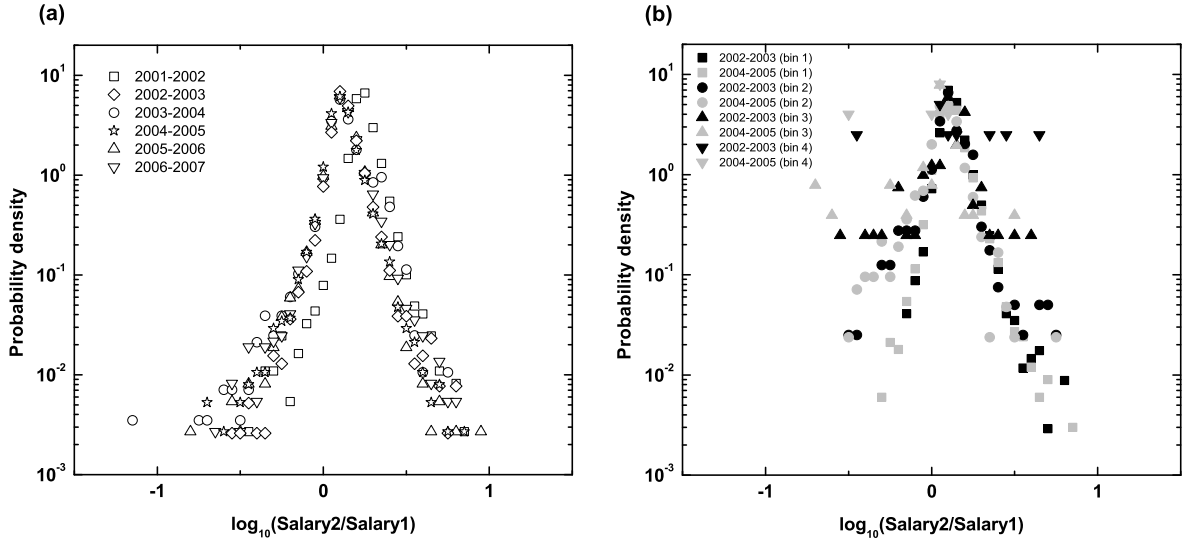}
  \end{center}
  \caption{Probability density for the logarithmic growth-rate. (a) Growth rates for the top income region, in consecutive years, excluding the years affected by the economic crisis (2007-2008 and 2008-2009). (b) Binned data points for years 2002-2003 and 2004-2005. The Pareto region is split into four intervals of equal length (bins) in the logarithmic income -- see text. The probability density functions are calculated for each interval. Results obtained in such manner are depicted with black color for years 2002-2003 and with gray for 2004-2005. For clearer visualization each bin is represented with a different symbol.}
  \label{fig7}
\end{figure}

One can also test the generality of the results obtained in \cite{aoyama_2003,fujiwara_2003}. In order to do that, we have computed the probability
density for the $log_{10}[W(t+\Delta t)/W(t)]$ logarithmic growth rate for several years and for several income intervals (bins).
Following the work of Fujiwara et al. \cite{fujiwara_2003}, we split the high income region $[W_{min},W_{max}]$ in four equal length intervals in the logarithmic income value 
($W \in [10^{(r_{min}+0.25(n-1)(r_{max}-r_{min})} ,  10^{(r_{min}+0.25n(r_{max}-r_{min})}] $, for $n\in\{1,2,3,4\}$ and $r_{max}=log_{10}(W_{max})$,  $r_{min}=log_{10}(W_{min})$).
Two separate and economically stable years (2002-2003) and (2004-2005) are analyzed in such manner. 
For year 2002 we obtained $r_{min}=3.4887$ and $r_{max}=5.04$, and for year 2004 we have $r_{min}=3.6386$ and $r_{max}=5.5243$.
The results obtained for the different years and different income intervals are presented on Figure \ref{fig7}.
The curves obtained for different years  -- Figure 7(a) -- collapse in an approximate manner and have a
similar shape with the ones drawn in \cite{aoyama_2003,fujiwara_2003}. The probability densities obtained in our studies are
however more extended in the negative region, and the peak is clearly shifted in the positive direction. The small decreases in salaries are less probable in Romania than
in case of Japan, but in Romania very large decreases are also present. This differences are a result of the fact that Japan has a developed and quite performing economy while Romania has a quickly changing and recently developing
market economy where salaries have an increasing trend and sometimes large fluctuations from year-to-year. 
It is important to observe also that the probability density plots obtained for different income intervals (bins) do not collapse, and
so our results do not confirm the generality of the results presented in \cite{fujiwara_2003}. In the case of the romanian economy the
growth rate is not independent of the income, as may be seen in Figure 7(b). This is again a major difference relative to the income dynamics of the Japanese employees, but we should recall that our study concerns just wages whereas the japanese one included other sources of income, like risky assets, which dominated in the total income.

\section{Conclusion}
In the present paper we investigated experimentally the workers' income distribution and its dynamics, using extended social security data for the Cluj district, Romania, for years 2001-2009. The difference between this study and previous real data income distribution studies performed in various countries of the World is
that our database allows a complete sampling in which each employee can be uniquely identified and his/her income can be followed in time. In such manner,
besides the characteristic distributions in the low and high income limits, extended information about the income dynamics can be also analyzed. To our knowledge, this is the first long time study of income mobility.
An interesting aspect of our data is
that it contains also information on two years (2007-2008) when a financial-economic crisis hit Romania. In this way, it is possible to identify the main effects of such a crisis
on the income distribution functions and income dynamics.

We have confirmed once again the validity of Pareto's principle for the upper $5\%$ income limit of the targeted society, obtaining an excellent scaling for four orders of magnitude in the rank, and a quite stable Pareto exponent $\alpha$ with less than $10\%$ yearly fluctuation
in the neighborhood of $\alpha=2.5$. The total income of each employee in the database, for the whole 9 years interval, exhibits also the Pareto scaling with
a similar exponent, $\alpha=2.69$. Although the proportionality constant for the power-law fitting varies in considerable manner for the studied
time-interval (suggesting large fluctuations in the average and maximal salaries) the Pareto exponent is unexpectedly stable.

In the low and medium income limit, we have confirmed the exponential-like shape of the income distribution function. The income temperatures
$T_r$, determined from the exponential fits, show a monotonic increase, except for 2008, where drastic austerity measures were
implemented as a result of the crisis, and the salaries of all state-employees were reduced by $25\%$.

The dynamical studies on the income of individuals in the high income limit revealed that the stability of the Pareto exponent over the 9 years is not a result
of a stable rank kept by the employees in the income top list, nor the consequence of the stability of the rank of top jobs. The Pareto limit is a highly dynamical region where both
the players and the jobs are constantly changing. The stability of the Pareto law and exponent should thus have a statistical explanation, in which fluctuations are the main
ingredient.

Following the dynamics of the income by plotting the changes in individual salary as a function of the previous salary,
we have confirmed a basic hypothesis used in current income models: the a multiplicative random growth of the income is a reasonable assumption for the richest sector of the population. In such
manner we gave experimental evidence for the assumptions of the models used in \cite{milakovic_2005,yakovenko_2009_springer}.
Our results do not confirm the fact that the growth-rate is independent of the income, as previously found for Japan \cite{aoyama_2003,fujiwara_2003}. It should be
noted, however, that in our case the income was exclusively salary, whereas in the Japanese study (based on income tax) income was dominated by profits from capital investments. Whether the income independence of growth rates is a direct consequence of the presence of these risky assets, or if major differences between the romanian and japanese economies are responsible for the two distinct behaviours, remains an open problem.

\section{Acknowledgments}

Financial support was provided by grant PN-II-ID-PCE-2011-3-0348  and from programs POSDRU 6/1.5/S/3 on Doctoral Studies.
N. Derzsy wishes to thank for the motivating working atmosphere during her research internship at Centro de F\'{\i}sica do Porto. We thank the Social Security Office Cluj for the provided data.

\bibliographystyle{model1-num-names}

\end{document}